\documentclass{aa} 
\usepackage{psfig}
\begin{document}
\def\lya{Ly$\alpha$}               %
\def\lyb{Ly$\beta$}                %
\def\Wr{w_\mathrm{r}}              %
\def\Wo{w_\mathrm{obs}}            %
\def\mr{m_\mathrm{R}}              %
\def\mb{m_\mathrm{B}}              %
\def\Mb{M_\mathrm{B}}              %
\def\mi{m_\mathrm{I}}              %
\def\Mi{M_\mathrm{I}}              %
\def\MAB{M_\mathrm{AB}(\mathrm{B})}%
\def\kms{~km~s$^{-1}$}             %
\def\cm2{~cm$^{-2}$}               %
\def\za{z_\mathrm{a}}              %
\def\zd{z_\mathrm{d}}              %
\def\ze{z_\mathrm{e}}              %
\def\zg{z_\mathrm{g}}              %
\def\h50{h_{50}^{-1}}              %
\def\hi{H\,{\sc i}}                %
\def\nhi{N(\mbox{H\,{\sc i}})}     %
\def\alii{Al\,{\sc ii}}            %
\def\caii{Ca\,{\sc ii}}            %
\def\civ{C\,{\sc iv}}              %
\def\crii{Cr\,{\sc ii}}            %
\def\feii{Fe\,{\sc ii}}            %
\def\mgi{Mg\,{\sc i}}              %
\def\mgii{Mg\,{\sc ii}}            %
\def\nv{N\,{\sc v}}                %
\def\ovi{O\,{\sc vi}}              %
\def\oii{[O\,{\sc ii}]}            %
\def\oiii{[O\,{\sc iii}]}          %
\def\znii{Zn\,{\sc ii}}            %
\thesaurus {03(11.17.1; 11.08.1; 11.09.4)} 
\title{\lya\ absorbers at $z\le 1$: HST-CFHT imaging and spectroscopy in the
field of 3C 286 
\thanks{Based on observations made with the NASA/ESA {\it Hubble Space 
Telescope}, obtained at the
Space Telescope Science Institute, which is operated by the Association of 
Universities for
Research in Astronomy, Inc., under NASA contract NAS 5-26555, and with the
Canada-France-Hawaii Telescope}}
\author{V. Le~Brun \inst{1} and J. Bergeron \inst{2,3}}
\institute{Laboratoire d'Astronomie Spatiale du C.N.R.S., B.P. 8, F-13376 
Marseille, France, lebrun@lasa13.astrsp-mrs.fr
\and European Southern Observatory, Karl-Schwarzschild-Stra\ss e 2, D-85748
Garching b. M\" unchen, Germany, jbergero@eso.org
\and Institut d'Astrophysique de Paris, CNRS, 98bis boulevard Arago, 
F-75014 Paris, France
}
\offprints{V. Le~Brun}
\date {Received 5 May 1997; accepted 20 Jan 1998}
\maketitle
\markboth{V. Le~Brun \& J. Bergeron: \lya\ absorbers toward 3C 286}{}
\begin{abstract}
We present further observational results  on our \lya -only absorber
identification project at low and intermediate redshifts. 
We combine CFHT/MOS imaging and spectroscopic observations of 25
galaxies in the field of the quasar 3C~286 ($\ze = 0.849$) with HST/FOS
spectrum of the quasar between 1600 and 3000~\AA \ and HST/WFPC2 high spatial
resolution imaging of some galaxies. 

Our results confirm our first conclusions on the nature of the intermediate
redshift \lya\ forest absorbers. A small fraction of them is tightly
linked to galaxies, and most probably arise in the external parts of
giant halos ($R\simeq200\h50$~kpc) around fairly luminous galaxies 
($M_\mathrm{AB}(\mathrm{B}) < -19$) of various spectral and
morphological types, as shown by HST imaging of these objects. Most of 
the absorbing clouds appear to be only 
``associated'' with galaxies, in the sense that they could trace the 
gaseous filaments or sheets of large-scale structures. Thus, as 
expected from a sample tracing large scale structures, these galaxies 
have various spectral types. 

We also find that groups of galaxies have always at least one 
associated \lya\ absorption line.

\end{abstract}   
%
\section{\label{intro}Introduction}
Observations with the Hubble Space Telescope (HST) have provided a large data
base of \lya\ absorption lines at low and intermediate redshifts, in
particular those from the HST Quasar Absorption Lines Key Project (Bahcall et
al. 1991, 1993, 1996). 
For the lower redshift \lya\ absorbers, several optical identification
searches have been conducted with ground-based telescopes to determine their 
nature, their cross-section and their link with  galaxy populations. Two 
complementary approaches have been pursued, one based on extensive local 
galaxy-redshift surveys, the other on large, homogeneous \lya\ absorption line 
samples.

The first approach was adopted by Morris et al. (1993), Stocke et al. (1995) 
and Shull et al. (1996) and requires to obtain  high-resolution and
signal-to-noise ratio spectra of nearby quasars located behind or within the 
regions sampled by the CfA redshift surveys (e.g. 3C~273, Mrk 501). There are 
usually only a few \lya\ absorption lines per sightline, furthermore with
very low rest-frame equivalent widths ($\Wr \simeq$ 20-300~m\AA), and the 
above authors have used  complete galaxy catalogs to characterize the 
locations of the  \lya\ absorbers relative to galaxies (including dwarf 
galaxies).
A crucial result is the  discovery of three \lya\ absorbers in
voids of  large-scale structures  and, for the most extreme case
(definite \lya\ absorption), the nearest known neighboring galaxy lies 
8.8 $\h50$~Mpc (where $h_{50}$ is the Hubble constant in units of 
50 km s$^{-1}$ Mpc$^{-1}$ and adopting $q_0=0$) 
away from the absorber (Stocke et al. 1995). This implies  that some 
\lya\ absorbers (mainly those of low column densities) are not tightly linked to
galaxies, or even to regions of matter overdensities. However, for the
majority of these low-redshift \lya\ absorbers (7/10 in the  Shull et
al.'s sample), there is a galaxy within $3.0\h50$~Mpc. 

In the other approach, studies of fields around quasars observed with the HST
Faint Object Spectrograph (FOS) aim
at identifying \lya\ absorbers at intermediate redshifts, $0.1<z<0.8$,
with larger
\lya\ equivalent widths, $\Wr \simeq 100-1000$~m\AA.  Lanzetta et al. (1995,  
hereafter referred to as LBTW) have found that the majority of the absorbers 
are directly associated with extended gaseous galactic halos, a result not 
confirmed by Le~Brun et al. (1996, hereafter referred to as LBB) who used a 
more homogeneous sample of \lya\ lines. 
 The main conclusion reached by LBB is consistent with the results obtained for
absorbers at lower redshift: the majority of \lya\ absorbers arise in gaseous 
regions spread along the large-scale structures of the galaxy distribution 
and only a small fraction of them trace extended galactic halos 
($R\simeq 200\h50$~kpc). Numerical simulations of the evolution of the gaseous
content of the Universe (Petitjean et al. 1995, M\"ucket et al. 1996)
are also consistent with this picture. 

The nature of these absorbers is more easily established at low redshift. If
the \lya\ absorbers are then shown to primarily trace large-scale structures,
their study over a large redshift range (up to 4-5) will provide information
on the evolution of large-scale structures and will thus constrain 
cosmological models.

In this paper, we present further observations from our identification 
programme  of  intermediate redshift \lya\ absorbers and report on our
results for the field around the quasar 3C~286. We have analyzed in details 
the UV spectrum of this quasar, for a study of the $\zd = 0.6921$
damped \lya\ system (Boiss\'e et al. 1998), as well as deep HST/WFPC2
images of the field surrounding the quasar. We have also obtained the 
spectra of 31 galaxies within 4\arcmin\ to the quasar sightline, successfully
measuring 25 redshifts. 

This data set increases the existing sample of galaxy/\lya -absorber pairs,
and helps to further constrain the
nature of the \lya\ forest. The data are presented in Sect.~\ref{data}, the
identified galaxies in Sect.~\ref{thegals}, and the discussion and conclusions
in Sect.~\ref{disc}.

\section{\label{data}Presentation of the data}
\subsection{\label{HSTspec}HST data}
\begin{figure}
\centerline{\psfig{figure=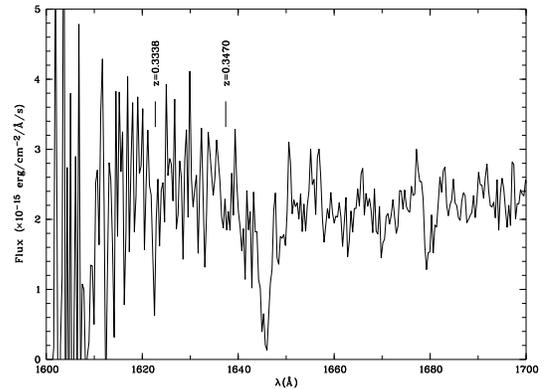,width=8cm,clip=t,angle=-90}}
\caption[]{\label{noisy} Part of the spectrum of 3C~286 showing the
lower reshift \lya\ absorption lines}
\end{figure}

\begin{table}
\caption{\label{abslines} List of the \lya\ systems detected in the spectrum
of 3C~286}
\begin{tabular}{rrrrrlr}
\hline\noalign{\smallskip}
$\lambda _\mathrm{obs}$ & $w_\mathrm{obs}$ & 
$\sigma\left(w_\mathrm{obs}\right)$ & $\Wr$ & $\lambda _\mathrm{r}$
& Ident. & $z_\mathrm{a}$\\
\hline\noalign{\smallskip}
1622.72 & 0.8  & 0.3 & 0.6  & 1215.67 & \lya & 0.3348 \\
1679.80 & 0.90 & 0.19 & 0.55 & 1025.72 & \lyb & 0.6377 \\
1711.37 & 0.82 & 0.18 & 0.58 & 1215.67 & \lya & 0.4078 \\
1739.49 & 0.86 & 0.18 & 0.60 & 1215.67 & \lya & 0.4309 \\
1753.86 & 0.19 & 0.14 & 0.11 & 1025.72 & \lyb & 0.7099 \\
1781.06 & 0.80 & 0.13 & 0.46 & 1025.72 & \lyb & 0.7364 \\
1799.71 & 1.01 & 0.14 & 0.68 & 1215.67 & \lya & 0.4804 \\
1813.43 & 0.93 & 0.14 & 0.62 & 1215.67 & \lya & 0.4917 \\
1911.47 & 0.86 & 0.11 & 0.55 & 1215.67 & \lya & 0.5724 \\
1928.22 & 0.80 & 0.11 & 0.50 & 1215.67 & \lya & 0.5861 \\
1990.53 & 1.45 & 0.13 & 0.88 & 1215.67 & \lya & 0.6374 \\
2078.20 & 0.55 & 0.12 & 0.32 & 1215.67 & \lya & 0.7095 \\
2111.73 & 1.03 & 0.10 & 0.59 & 1215.67 & \lya & 0.7371 \\
2143.86 & 0.30 & 0.08 & 0.17 & 1215.67 & \lya & 0.7635 \\
2149.72 & 0.83 & 0.10 & 0.47 & 1215.67 & \lya & 0.7683 \\ 
2152.64 & 0.34 & 0.09 & 0.19 & 1215.67 & \lya & 0.7707 \\
2179.61 & 0.28 & 0.07 & 0.16 & 1215.67 & \lya & 0.7929 \\
2183.11 & 0.52 & 0.08 & 0.29 & 1215.67 & \lya & 0.7958 \\
2224.00 & 0.19 & 0.06 & 0.10 & 1215.67 & \lya & 0.8294 \\
\noalign{\medskip}\hline
\end{tabular}
\smallskip\par
\end{table}
The UV spectrum of 3C~286 ($\ze=0.849$) was obtained in February 1995 with 
the HST/FOS and the grisms G190H-G270H (wavelength 
coverage 1620-3270~\AA), as part of a programme devoted to a study of 
intermediate redshift damped \lya\ absorbers. The details of the data 
reduction and analysis are presented in Boiss\'e et al. (1998). A systematic 
search of \lya\ lines was made in the wavelength range 
1650-2240~\AA, the region between 1620 and 1650~\AA\ being of lower 
signal-to-noise ratio.The line list is 
given in Table~\ref{abslines}, excluding the features associated 
with the damped \lya\ system (DLAS). Boiss\'e et al. have
selected  lines detected at the $4.5\sigma$ level, whereas we have 
included a few additional candidate \lya\ features which are 
significant only at the $3\sigma$ level. The first line of the 
list ($\lambda_\mathrm{obs} = 1622.7$~\AA), as well as a line at
1637.4~\AA, were search for because we have detected galaxies at
these redshifts in the quasar field. These absorption features
have detection levels of respectively 
$2.6$ and $1.3\sigma$ (Fig.~\ref{noisy}). Thus, 
only the first one will be used in the subsequent analysis.
Since Boiss\'e et al. (1998) paid special 
attention to identify all 
weak metal lines at $\zd=0.6921$, which is the only strong metal-rich system
in the spectrum of 3C~286, it is then most likely that the sample 
presented in Table~\ref{abslines} is not contaminated by metal lines.

We have detected a total of 10 \lya\ absorption lines 
with a limiting rest equivalent width of 0.24~\AA, in the  redshift interval 
$zz = 0.357,0.842$, thus a density of 20.9 lines per unit redshift. This 
value is in good agreement (within $1\sigma$) with that derived from 
the HST QSO absorption lines Key Project for the same limiting
rest equivalent width (Bahcall et al. 1996):
\begin{equation}
{\mathrm{d}N\over \mathrm{d}z}=23\pm 4 \mbox{ at } \left<z\right>=0.6\,.
\end{equation}

HST images were obtained for the DLAS programme, and the data are
fully described in Le~Brun et al. (1997). These images have
a $5\sigma$ limiting magnitude $m_{702}=26.15$.

In addition, we have retrieved from the HST data bank (dataset root Z2P4010) 
a higher resolution ($R=20000$) spectrum of the quasar obtained by 
Briggs et al.  with the Goddard High Resolution Spectrograph and the 
grating G160M (wavelength range 1540-1580~\AA) in June 1995. Except 
for the lines of the Lyman series at $\zd=0.692$, no other line is 
present in this spectrum. This sets limits on the \civ\ absorption 
associated with some galaxies
identified in the field of 3C~286 (see Sect.~\ref{thegals}). 
\begin{figure*}
\centerline{\psfig{figure=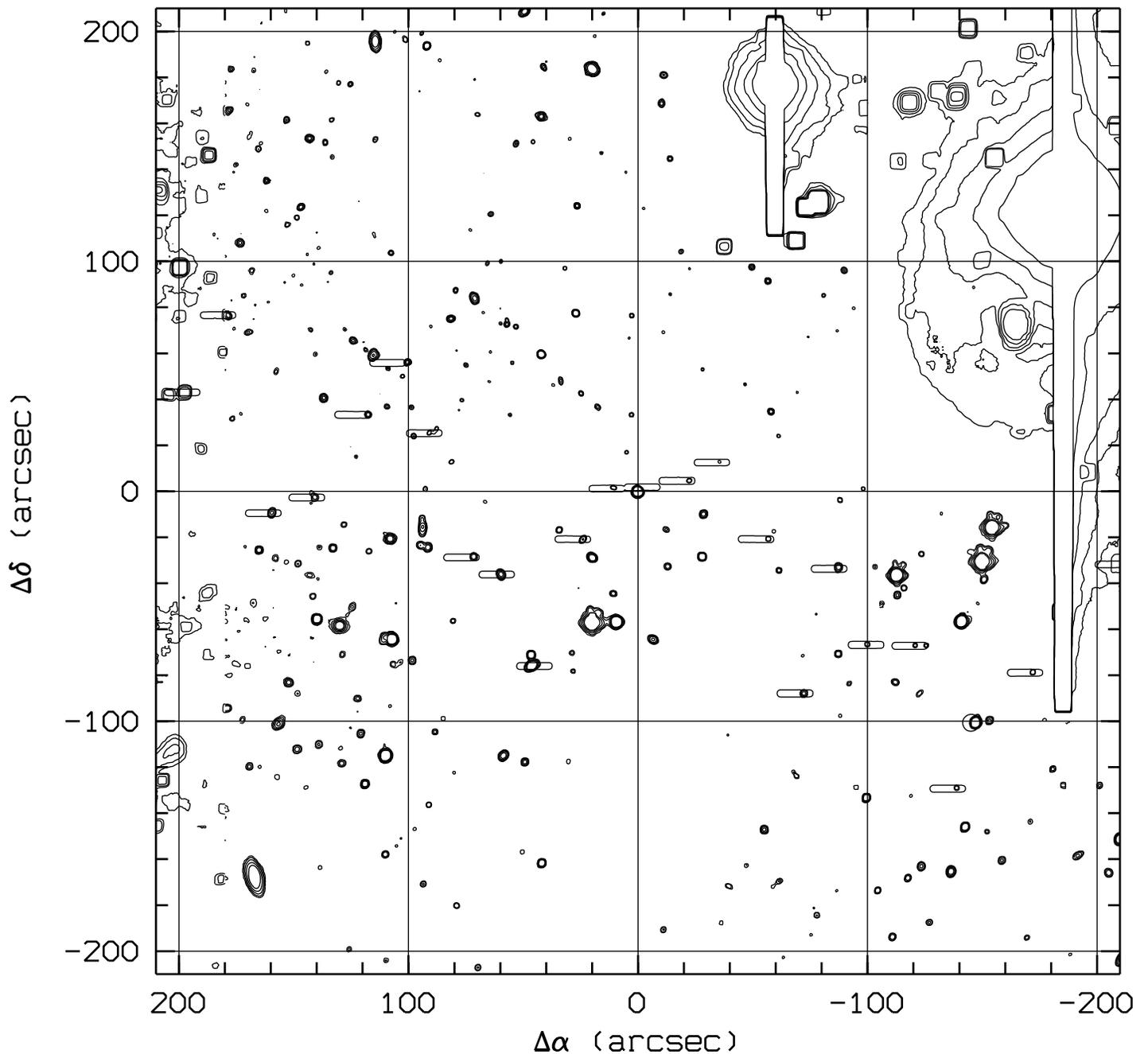,height=18cm,clip=t,angle=0}}
\caption[]{\label{field} Field of the CFHT/MOS around 3C~286. The 
galaxies spectroscopically observed with MOS are encircled}
\end{figure*}
\subsection{Galaxy population of the field (MOS imaging)}
These data were obtained in May 1993 with the CFHT Multi 
Object Spectrograph (MOS). The data reduction and analysis techniques 
are as described in LBB. Images of the overall field were also obtained 
with the MOS, with a plate scale of 0.314\arcsec\ per pixel, down to a
$5\sigma$ limiting magnitude $\mr=23$. The field is reduced
from the original 10\arcmin\ MOS field of view by vignetting and
shift-and-add procedure to a $8.3\arcmin\times8.3\arcmin$ 
field of view.

The detection and photometry of the galaxies in the field were made 
with the software SExtractor package (Bertin \& Arnouts 1996). Our 
goal (not yet reached) for the identification survey of the galaxies 
associated in any way to \lya\ absorbing  clouds is to obtain a 
complete galaxy sample down to a limiting magnitude $\mr = 22.5$ and 
with an angular impact parameter $\theta \le 3.5\arcmin$. The field is 
shown in Fig.~\ref{field}; the contours of the MOS slitlets show the 
galaxies for which spectra have been obtained. There are 127
galaxies in the above defined ranges of magnitude and impact 
parameter, which distribution versus magnitude is given in Column~2 of 
Table~\ref{galpop}. 
\begin{table}
\caption{\label{galpop} Galaxy population of the field and completeness
of the spectroscopic sample}
\begin{tabular}{rrrr}
\noalign{\smallskip}
Magnitude range & 
N$_\mathrm{gal}$ & 
N$_\mathrm{z}$   & 
Comp.$^\mathrm{c}$ \\
                &
$\left(\theta\le 3.5\arcmin\right)^\mathrm{a}$ &
$\left(\theta\le 3.5\arcmin\right)^\mathrm{b}$ &
                   \\
\noalign{\smallskip\hrule\medskip}
$ \ \ \ \mr \le 17.5$&  0 &  0 &  -    \\
$17.5 < \mr \le 18.5$&  1 &  1 & 100\% \\   
$18.5 < \mr \le 19.5$&  3 &  0 &  25\% \\
$19.5 < \mr \le 20.5$& 13 &  2 &  18\% \\
$20.5 < \mr \le 21.5$& 34 &  8 &  22\% \\
$21.5 < \mr \le 22.5$& 76 &  8 &  15\% \\
\noalign{\smallskip}\hline
\end{tabular}
\smallskip\par
$^\mathrm{a}$ Number of galaxies in each magnitude bin\noindent \\
$^\mathrm{b}$ Number of galaxies with measured redshift   \noindent \\
$^\mathrm{c}$ Completeness with regard  to the cumulated magnitude 
distribution 
\end{table}
\subsection{\label{spectro}MOS spectroscopic data}
We have obtained 31 spectra over the MOS field, from which we have 
derived 25 redshifts (see Table~\ref{objects}), corresponding to a success rate of 
about 81\%, close to that reached by LBB. The spectral r\'esolution
is $R=1500$, and the error on the 
redshift estimate is about $\Delta z = 0.0004$, due to the
modest signal-to-noise ratio of the spectra. In the 
3.5\arcmin \ radius circle, we got redshifts for 19 galaxies, and the 
completeness of the spectroscopic sample is given in
Column~3 of Table~\ref{galpop}, 
together with the number of galaxies in each magnitude bin (Column~2).
The absolute magnitudes are determined using a $k$-correction which is 
estimated from a comparison between two spectra: 
firstly, the region of the galaxy rest-frame spectrum corresponding to the 
observed $r$-band wavelength range and, secondly, the B-band wavelength region of 
the template spectrum of a galaxy of same morphological type (as determined by 
the Balmer+\caii\ break) given by Coleman et al. (1980). In addition, the shift
from Vega-type to AB magnitudes is given by $M_\mathrm{AB}(\mathrm{B}) = 
M(\mathrm{B}) - 0.17$.

From the 25 identified galaxies, 12 are within 750~\kms \ from a \lya\
absorption line (this velocity criterion was defined in LBB, and is adequate
to test the  large-scale structure membership), and two show no associated 
\lya\ absorption, one of them being at the quasar redshift. Two additional 
galaxies are at redshifts for which there is no observation in the wavelength 
range of  the expected associated \lya\ absorption, while for seven others the 
expected \lya\ line would be blueward of the Lyman discontinuity  at
$\zd=0.692$. Finally, two objects are background galaxies.  
\begin{table*}
\caption{\label{objects} Results of the spectroscopic observations}
\begin{tabular}{rrrrrrrrr}
\hline\noalign{\smallskip}
Slit & $\Delta x$(\arcsec) & $\Delta y$(\arcsec) & $\theta$(\arcsec) & 
$D$($\h50$ kpc) & $\zg$ & $\za$ & $\mr$  & $\MAB$ \\
\hline\noalign{\smallskip}
QSO&   0.0 &   0.0 &   0.0 &    0 & 0.849  &        & 17.04 & $-$26.45 \\
 1 &       &       &       &        & indef  &      &       &        \\
 2 & 251.1 & 113.1 & 275.4 & 1983 & 0.4070 & 0.4078 & 20.65 & $-$21.68 \\
 3 & 232.7 &  96.8 & 252.0 & 1642 & 0.3457 & 0.3470$^\mathrm{a}$ & 22.81 & $-$19.18 \\
 4 & 232.1 & $-$40.0 & 235.5 & 1883 & 0.4900 & 0.4917 & 23.25 & $-$19.71 \\
 5 & 209.7 & $-$31.5 & 212.0 & 1301 & 0.3152 &\lya\ wavelength range not observed
                                                    & 21.22 & $-$20.51 \\
 6 & 172.1 & $-$78.7 & 189.2 & 1957 & 0.8599 & background galaxy
                                                    & 22.18 & $-$22.57 \\
 8 & 138.8 &$-$129.0 & 189.5 & 2001 & 0.9090 & background galaxy
                                                    & 21.97 & $-$22.97 \\
9a & 125.6 & $-$67.1 & 142.4 &      & indef. &        & 22.48 &        \\
9b & 120.8 & $-$67.0 & 138.1 &  876 & 0.3316 & 0.3348 & 22.32 & $-$19.58 \\
10 &  99.9 & $-$66.5 & 120.0 & 1093 & 0.6350 & 0.6374 & 21.95 & $-$22.40 \\
11 &  87.4 & $-$33.2 &  93.5 &  819 & 0.5860 & 0.5860 & 20.65 & $-$23.23 \\
12 &  72.3 & $-$87.9 & 113.8 &  692 & 0.3107 & \lya\ wavelength range not observed
                                                    & 20.63 & $-$21.08 \\
13 &  56.8 & $-$20.7 &  60.5 &      & indef  &        & 22.13 &        \\
14 &  35.5 &  12.8 &  37.7 &      & indef  &        & 22.79 &        \\
15 &  22.3 &   4.6 &  22.8 &  234 & 0.8474 & no associated \lya\ absorption line 
                                                    & 22.09 & $-$22.62 \\
16 &   8.1 &   2.3 &   8.4 &      & indef  &        & 23.81 &        \\
17 & $-$10.7 &   1.6 &  10.8 &   68 & 0.3338 & 0.3348 & 21.98 & $-$19.95 \\
18 & $-$24.0 & $-$20.9 &  31.8 &  166 & 0.2488 & blueward to the $\zd=0.6921$ LL 
                                                    & 21.16 & $-$19.97 \\
19 & $-$46.2 & $-$75.8 &  88.7 &  461 & 0.2475 & blueward to the $\zd=0.6921$ LL 
                                                    & 18.44 & $-$22.39 \\
20 & $-$59.7 & $-$36.2 &  69.8 &   88 & 0.0463 & blueward to the $\zd=0.6921$ LL 
                                                    & 20.12 & $-$17.38 \\
21 & $-$71.5 & $-$28.4 &  77.0 &  750 & 0.7394 & 0.7371 & 21.23 & $-$23.05 \\
22 & $-$90.9 &  25.5 &  94.4 &      & indef  &        & 22.42 &        \\
23 &$-$100.4 &  56.0 & 115.0 &  859 & 0.4334 & 0.4309 & 21.19 & $-$21.45 \\
24 &$-$117.5 &  33.4 & 122.2 &  977 & 0.4895 & 0.4917 & 21.36 & $-$21.43 \\
25 &$-$140.7 &  $-$2.6 & 140.8 & 1084 & 0.4575 & no associated \lya\ absorption line 
                                                    & 21.67 & $-$21.11 \\
26 &$-$159.5 &  $-$9.3 & 159.8 &  837 & 0.25   & blueward to the $\zd=0.6921$ LL 
                                                    & 21.14 & $-$20.01 \\
27 &$-$178.5 &  76.5 & 194.2 & 1453 & 0.4350 & 0.4309 & 21.41 & $-$21.24 \\
28 &$-$197.4 &  43.0 & 202.0 & 1615 & 0.4900 & 0.4917 & 20.35 & $-$22.61 \\
29 &$-$207.9 & 130.6 & 245.6 & 1170 & 0.2193 & blueward to the $\zd=0.6921$ LL 
                                                    & 19.37 & $-$21.43 \\
30 &$-$233.5 &  80.4 & 246.9 & 1286 & 0.2479 & blueward to the $\zd=0.6921$ LL 
                                                    & 20.08 & $-$21.05 \\
31 &$-$245.2 &  75.8 & 256.6 &  305 & 0.0435 & blueward to the $\zd=0.6921$ LL 
                                                    & 15.71 & $-$20.87 \\
\hline\noalign{\smallskip}
\end{tabular}
\smallskip\par
$^\mathrm{a}$ This absorber will not be included in the subsequent analysis 
\end{table*}
\section{\label{thegals}Characteristics of the identified galaxies}
\subsection{The lower redshift objects}
As mentioned above, the Lyman discontinuity associated with the DLAS prevents
to detect any \lya\ line at $\za\le 0.27$. In addition, the observed
wavelength range restricts the detection of \lya\ lines to $z\ge 0.33$. 
Therefore, we could only  search for possibly associated metal
lines for the objects described in this section.

We note that there is neither a  low redshift ($z<0.03$) \civ\ absorption 
 nor a local galaxy at an impact parameter smaller than $D\sim 100\h50$~kpc.
For the latter search, we have used the Simbad database\footnote{The Simbad 
database is operated at CDS, 
Strasbourg, France} to list the local galaxies within 1 degree from the
 quasar. There are 16 such galaxies, but none of them 
has an impact parameter small enough to expect associated \civ\ absorption. 
\begin{figure}
\centerline{\psfig{figure=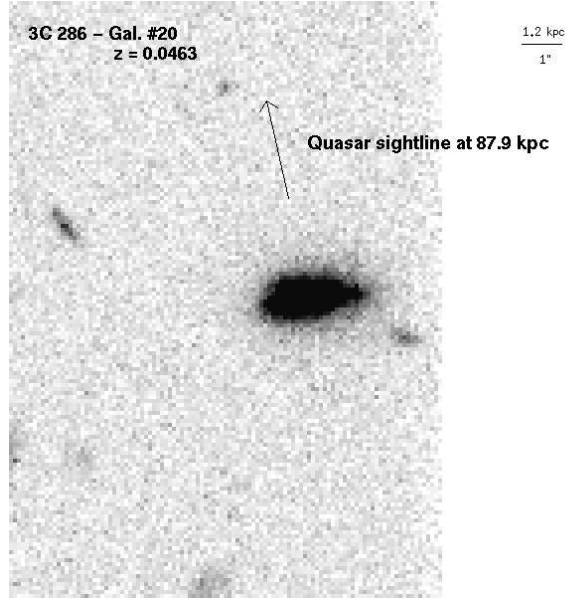,height=8cm,clip=t}}
\caption[]{\label{gal20} HST/WFPC2 image of the dwarf galaxy \#20.}
\end{figure}
%
\subsubsection{The group at $z=0.045$}
This group has two identified members, \#20 and \#31, with impact parameters 
$D=87.9$ and $304.8 \h50$~kpc respectively, and separated by 800\kms.
The closest galaxy is very faint, $\MAB = -17.38$, with a peak surface brightness 
$\mu_\mathrm{702,max}=21.2$, and is detected in our HST 
images  (see Fig.~\ref{gal20}). Its linear extent is $4.8\times3.2\h50$~kpc. 
This  object is similar to the well known dwarf 
galaxy detected by Steidel et al. (1995) at $z=0.072$, close to the sightline
to PKS~0454+039, with $\Mb=-17.2$, $\mu_\mathrm{702,max}=20.4$, and a linear
extend of $5.1\times2.8\h50$~kpc. The impact parameters of these two dwarf 
galaxies differ by one order of magnitude ($D=7.6$ and 87.9 $\h50$~kpc for the 
galaxies in the fields of  PKS~0454+039 and  3C~286 respectively),
which is consistent with the detected associated \mgii\ absorption in the spectrum 
of  PKS~0454+039 and the lack of such an absorption  in  
 3C~286 (Boiss\'e et al. 1998). The associated \civ\ doublet is
unobservable for PKS~0454+039, due to a LLS at $z=0.8596$, and
is expected in a very noisy region of our HST/G190H spectrum of 3C~286, 
thus preventing detection. 
\subsubsection{The group at $z=0.248$}
This group has an average redshift $z=0.248$ and includes at least 3 
members (galaxies \#18, \#19 and \#30), spanning 310\kms. The impact
parameters range from 166 to $1286\h50$~kpc, and the absolute magnitudes
from $-20.0$ to $-22.4$. There is no detection of the associated \civ\ 
lines, expected at 1932.3 and 1935.6~\AA, with a limiting rest equivalent
width of 0.3~\AA.  

We note the presence of another fairly bright ($m_\mathrm{V}=17.8$) quasar,
1E~1328.5+3135, at a redshift $z=0.241$, or 1700\kms\ from this group,  
50\arcmin\ north to 3C~286. This angular separation corresponds to 
$15\h50$~Mpc at $z=0.241$. This quasar may then belong to the same 
large-scale structure than the above group of galaxies. 
\subsubsection{The group at $z=0.313$}
There are two galaxies (\#5 and \#12), separated by 1030\kms, which may be
part  of  a group. They have impact parameters $D=1300\h50$ and $692\h50$~kpc,
and both have about $L^\star$ luminosities. 
\subsection{The group at $z=0.333$}
\begin{figure}
\centerline{\psfig{figure=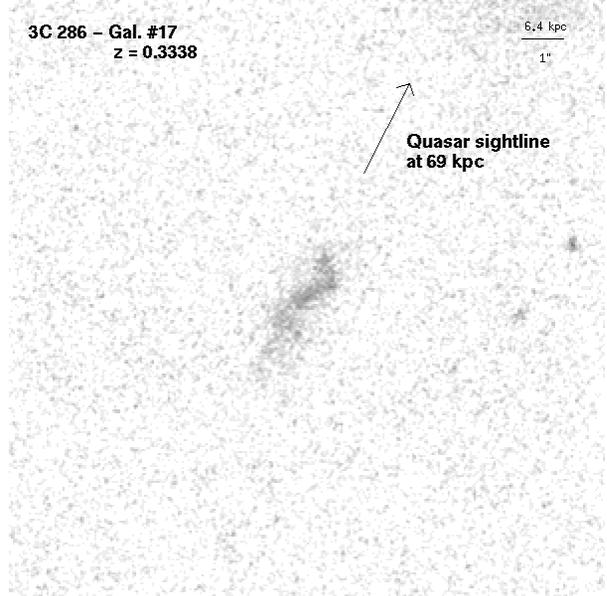,height=8cm,clip=t}}
\caption[]{\label{gal17} HST/WFPC2 image of galaxy \#17.}
\end{figure}

This group has two identified members (galaxies \#9b and
\#17), at an average redshift $z=0.333$, and with a linear projected
separation of $300\h50$~kpc. There is a  \lya\ absorption line at 
$\za = 0.3348$, which we assume to be associated with the detected 
galaxy  of  
 smallest impact parameter, \#17 at $D=68.8\h50$~kpc, with a velocity
difference $\Delta v = 220$\kms . The HST image of 
this galaxy ($M_\mathrm{AB}(\mathrm{B})=-19.95$) is shown in 
Fig.~\ref{gal17}. It is a very
diffuse amorphous object, which does not clearly match any type of the
Hubble sequence, as seen in the HST Medium Deep Survey (Abraham et al. 
1996), or in the Hubble Deep Field (Van den Bergh et al. 1996). The 
comparison
of the magnitudes measured from CFHT and HST data shows that there
is no dominant very-low surface brightness component in this object 
(that would have been detected in CFHT but not in HST images). 
The spectrum of this galaxy shows strong \oii\ ($\Wr=26$~\AA) and 
fainter emission lines of \oiii\ ($\Wr=7.2$~\AA),  H$\alpha$ 
($\Wr=12$~\AA) and H$\beta$ ($\Wr=7$~\AA). 
The galaxy impact parameter is slightly larger than the value derived 
from the $(M,R)$ scaling law by Guillemin \& Bergeron (1997)
for the size of the \mgii \ absorbing halo, 
$R = 90 (L_B/L_B^\star)^{0.28} \h50$ kpc. This is consistent with the 
lack of detection of an associated \mgii \ absorption in the spectrum 
of 3C~286 by Aldcroft et al. (1994).
The  \civ\ absorption doublet is expected at  2064.99 and 2068.41~\AA, thus  
in the red wing of the damped \lya\ absorption line at $\zd=0.6921$. 
Higher resolution and signal-to-noise observations are required to 
better define the profile of the damped \lya\ line and search for weak 
\civ\ absorption. Finally, the \nv\ doublet, expected at 1652.34 and 
1657.64~\AA, is not detected at a $3\sigma$ rest equivalent width 
limit of $0.43$~\AA. 

There is one other galaxy in the field (\#3) at $z=0.3457$, which is 
likely to belong to an adjacent large-scale structure. An associated 
\lya\ absorption line may be present (see Sect.~\ref{HSTspec}) at 
$z=0.3470$ or $\Delta v = 290$\kms\ from the galaxy redshift. The 
magnitude and impact parameter of galaxy \#3  are $D=1642.0\h50$~kpc 
and $\MAB = -19.18$. However, due to the uncertainty on the reality of 
the \lya\ absorption line, this galaxy-absorber association will not 
be used in the subsequent analysis. 
\subsection{The group at $z=0.434$}
This group comprises two identified galaxies (\#23 and \#27), with large 
impact parameters $D=850$ and $1450\h50$~kpc respectively, with a projected 
separation of $585\h50$~kpc. Both galaxies are bright, $\MAB = -21.45$ and
$-21.24$, and they are separated by only 330\kms. There is 
a fairly strong \lya\ absorber associated with this group at $\za=0.4309$ with 
$\Wr = 0.42$~\AA , thus separated by 520 and 850\kms\ from galaxies \#23 and
\#27 respectively. The associated \civ\ doublet is not detected at 
a $3\sigma$ limiting rest equivalent width of 0.18~\AA\ for the \civ1548 line.
\subsection{The group at $z=0.490$}
\begin{figure}
\centerline{\psfig{figure=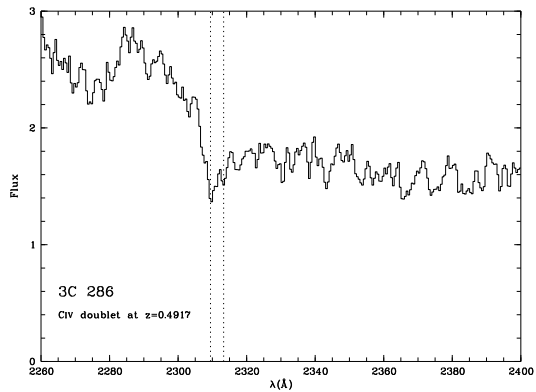,angle=-90,width=8cm,clip=t}}
\caption[]{\label{civ}Part of the HST/FOS spectrum showing the \civ\ doublet at $z=0.4917$}
\end{figure}

This group has 3 identified members at $z=0.4900$ within 100\kms 
(\#4, \#24 and \#28, $\MAB = -19.71, -21.43$ and $-22.61$ 
respectively). These galaxies cover a large range of spectral
type, with either emission lines  (galaxy \#4), absorption lines (\#24), or 
 both absorption and emission lines (\#28). Their impact parameters range 
from 975 to $1880\h50$~kpc and  the total projected extent of this group 
is larger than 3.5$\h50$~Mpc. There is an associated  \lya\ absorber  at 
$\za=$0.4917 with  $\Wr =0.62$~\AA, which is outside the velocity range covered by the
galaxies by 350\kms. 
There is a weak, but clearly present, associated \civ\ absorption doublet  in
the quasar UV spectrum (see Fig.~\ref{civ}), with 
$\Wr(\mbox{\civ}1548,1550)=0.24,0.12$~\AA, but the \nv\ doublet is not 
detected at a $3\sigma$ rest  equivalent width limit of 0.29~\AA.   
It is likely that this absorption arises in a yet non-identified galaxy which 
lies closer to the quasar sightline, since \civ\ absorption is expected to 
occur in the external parts of galactic halos. There are 10 unidentified 
galaxies within 40\arcsec\ to the quasar sightline, or $300\h50$~kpc at 
$z=0.4900$, which could be responsible for this \civ\ absorption. An
alternative hypothesis is that this \civ\ absorption could be due to
intra-group (or cluster) gas. The observation of higher ionization
level ions (like \nv\ or \ovi ) could help discriminating these two
origins.
\subsection{Isolated foreground galaxies}
\subsubsection{The galaxies at $z=0.4070$ and $0.4575$}
These galaxies, \#2 and \#25, have redshifts $\zg = 0.4070$
and 0.4575 respectively. Galaxy \#2 ($\MAB = -21.68$) has an associated \lya\ 
absorber, at $\za = 0.4078$, or a velocity difference of 170~\kms , for an 
impact parameter of $2.0\h50$~Mpc. Its spectrum shows strong H \& K 
absorption, as well strong G band and Mg\,{\sc i} absorptions.

On the contrary,
galaxy \#25 ($\MAB = -21.11$), with impact parameter $D=1.08\h50$~Mpc, does 
not have any associated \lya\ absorption, down to a $3\sigma$ limiting rest 
equivalent width of 0.24~\AA. Its spectrum shows \oii 3727 emission, as well 
as H \& K absorption lines. 
\subsubsection{The galaxies at $z=0.5860$ and $z=0.6350$}
Each of these two galaxies (\#11 and \#10 respectively, $\MAB = -23.23$ and
$-22.40$) has an associated \lya\ absorbing cloud, with
impact parameters $D=819.1$ and $1093.2\h50$~kpc and relative velocity
$\Delta v = 0$ and $440$\kms , for $z=0.5860$ and $z=0.6350$ respectively. 
Both galaxies have absorption line spectra, in addition to an \oii 3727
emission for galaxy \#11.
\begin{figure}
\centerline{\psfig{figure=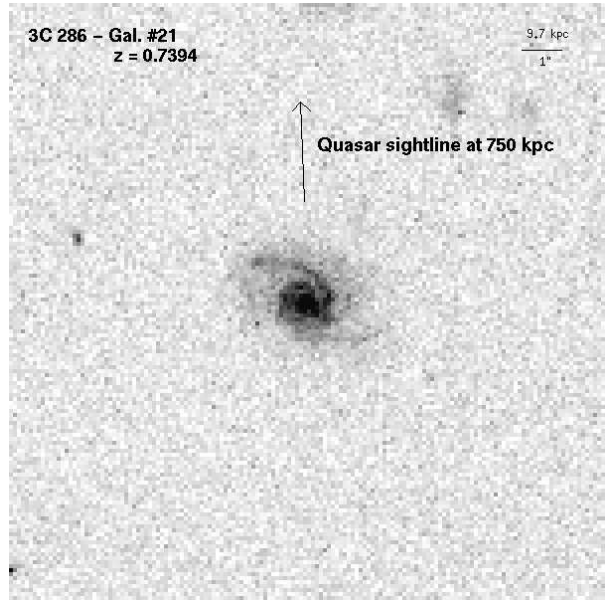,height=8cm,clip=t}}
\caption[]{\label{gal21} HST/WFPC2 image of the galaxy \#21, which shows 
associated \lya\ absorption.}
\end{figure}
%
\subsubsection{The galaxy at $z=0.7394$}
Based on the presence of a single emission line in the spectrum, identified
as \oii 3727 , this galaxy (\#21) has a redshift $\zg = 0.7394$,
or 400\kms\
away from the $\za=0.7371$ absorber. This galaxy is in
the observed field of the WFPC2 that we obtained for
the identification of the $\zd=0.692$ DLAS absorber. The 3600~s exposure of
the 15\arcsec\ field around the galaxy is shown on Fig.~\ref{gal21}. This
object appears to be a nearly face-on bright ($\MAB = -23.02$) spiral galaxy, 
with two well-developed arms extending over 5.4\arcsec, or $53.0\h50$~kpc. 
The spectrum of this galaxy shows a faint \oii\ emission ($\Wr = 9$\AA), and 
no \oiii\ emission ($\Wr \le 7$~\AA), nor absorption lines. 
\subsection{The galaxy at $z=0.8474\simeq\ze$}
Object \#15 is the only identified bright ($\MAB = -22.62$) galaxy with impact 
parameter smaller than $250\h50$~kpc ($D=234.4\h50$~kpc) that does not show 
any associated \lya\ absorption line in a sample of 43 (LBB) + 12 (this work) 
galaxies suitable for the study of the galaxy-\lya-absorber relation. As 
this object is only within 300\kms\ of the quasar redshift, it could  
be located behind the quasar. Alternatively, a gaseous halo associated 
with this galaxy could be highly ionized by the quasar UV flux 
 and thus contain only a very small amount of neutral hydrogen.    
Therefore, this galaxy, as well as any other galaxy within 3000\kms\ of the quasar
redshift, is excluded from our sample.
\section{\label{disc}Discussion on the \lya\ forest absorbers}
\begin{table}
\caption{\label{sample}The sample of identified \lya\ absorbers in the field
of 3C 286}
\begin{tabular}{lrrlrlr}
\hline\noalign{\smallskip}
$\za$  & $\Wr$ & Gal. & $\zg$  & $D$        & $\MAB$   & $\Delta v$ \\
       & \AA   &  \#  &        & $\h50$~kpc &          & km s$^{-1}$\\
\hline\noalign{\smallskip} 
0.3348 &  0.6  & 17   & 0.3338 &   68.8     & $-$19.95 & $-$220 \\
0.4078 &  0.58 & 2    & 0.4070 & 1983.0     & $-$21.68 & $-$170 \\
0.4309 &  0.60 & 23   & 0.4334 &  858.7     & $-$21.45 &    520 \\
0.4917 &  0.62 & 24   & 0.4895 &  976.8     & $-$21.43 & $-$440 \\
0.5860 &  0.50 & 11   & 0.5860 &  819.1     & $-$22.23 &     20 \\
0.6374 &  0.88 & 10   & 0.6350 & 1093.2     & $-$21.40 & $-$440 \\
0.7394 &  0.59 & 21   & 0.7371 &  749.6     & $-$23.05 & $-$400 \\
\multicolumn{2}{r}{$\le$0.36}& 25   & 0.4575 & 1083.9     & $-$21.11 &        \\
\hline\noalign{\smallskip} 
\end{tabular}
\end{table} 
\subsection{Overall properties of the galaxy-\lya\ absorber associations}
Table~\ref{sample} lists the new sample of seven \lya\ absorber-galaxy 
associations, and one galaxy without associated \lya\ absorption
detected in the field toward 3C~286. As in LBB, when several galaxies 
could be associated with the same \lya\ absorber, we have
only considered the closest one. This criterion had to be applied in four
cases, therefore excluding the galaxies \#9b in the $z=0.332$ group, \#27 in 
the $z=0.434$ group as well as \#4 and \#28 in the $z=0.490$ group. By 
comparison, we had to apply this criterion for only two galaxies in the larger 
sample of LBB. 
This may be linked with galaxy clustering (see Sect.~\ref{groups} for 
further discussion about this point) along this particular
sightline and supports LBB suggestion that there is not a one-to-one 
correspondance  between \lya\ absorbers and galaxies, contrary to what was 
found for metal-rich absorbers. 

\begin{figure}
\centerline{\psfig{figure=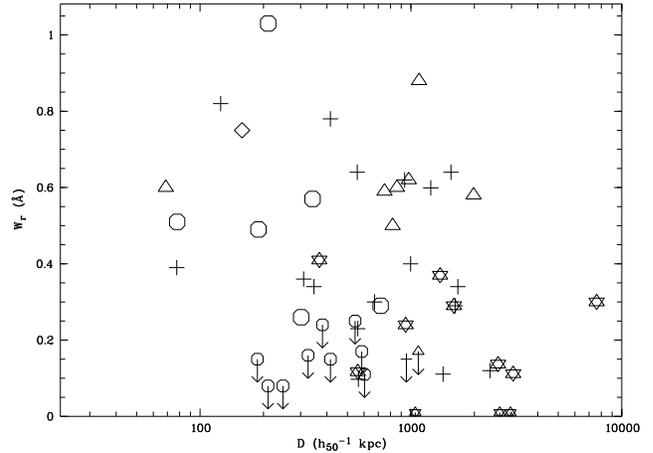,height=6.5cm,clip=t,angle=-90}}
\caption{\label{D_W} \lya\ line rest equivalent width plotted against galaxy 
impact parameter. Triangles: this work. Crosses: LBB. Circles: LBTW. Stars:
Morris et al. (1993). Diamond: Schneider et al. (1992)}
\end{figure}

\begin{figure}
\centerline{\psfig{figure=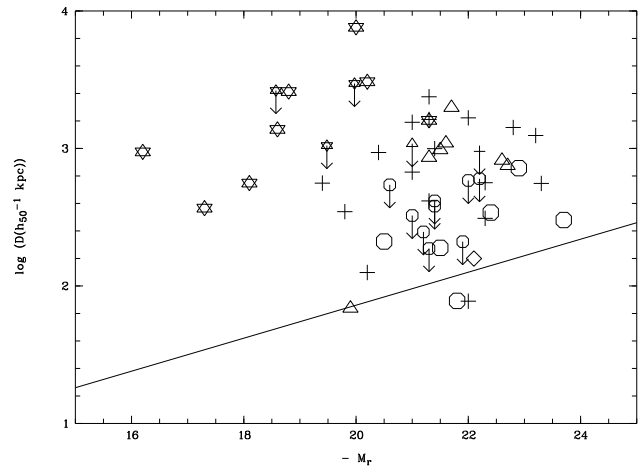,height=6.5cm,clip=t,angle=-90}}
\caption{\label{M_D} Galaxy impact parameter plotted against galaxy
luminosity. Same symbols as in Fig.~\ref{D_W}. The solid line represents the
$(\MAB,D)$ scaling law for \mgii\ absorbers, as given in Guillemin \& Bergeron
(1997)}
\end{figure}

Our new sample has been added to that studied in LBB (32 \lya\ absorber-galaxy 
associations and 11 galaxies without associated \lya\ absorption) which 
included results from other published \lya\ absorber surveys. The latter 
sample has also been modified in the following way:
\begin{itemize}
\item the $\zg=0.5200$ galaxy, which has no associated \lya\ absorption, in 
the field of  US~1867 (LBTW) has been removed, since it lies only within 
3000\kms\ from the quasar;
\item three galaxies without associated \lya\ absorption from the sample of 
Morris et al. (1993) have been included; they have redshifts $\zg=0.04375$, 
0.07789 and 0.10343, and impact parameters $D=2.64, 1.05$ and $2.97\h50$~Mpc 
respectively. 
\end{itemize}
The overall sample now contains 53 galaxies, of which 39 \lya\ absorber-galaxy
 associations (galaxies at less than 750\kms\ from a \lya\ absorbing
cloud) and 14 galaxies without associated \lya\ absorption. It 
is now well known that both damped \lya\ absorption systems and strong \mgii\ 
absorption systems\footnote{that is with $\Wr(2796) \ge 0.3$~\AA , these systems
are systematically associated with partial or total Lyman Limit Systems} are 
strongly linked to galaxies, with one-to-one association (Bergeron \& Boiss\'e 
1991, Steidel 1995, Le Brun et al. 1997), and with an anti-correlation between 
the impact parameter and the equivalent width of \mgii\ (and thus likely \lya) 
lines. Thus, we have included in our sample only 
\lya\ lines with $\tau_{912} < 1$. It appears in fact that these two sub classes
of absorbers do not have the same relation to galaxies, and we believe 
that including stronger \lya\ absorption lines in the sample of the 
so-called '\lya\
forest lines' would bias the correlation analysis.
Thus, when considering only 'faint' \lya\ lines, we still do not find any 
correlation  between  either the impact parameter and the \lya\ 
rest equivalent width, nor between the galaxy luminosity and the impact 
parameter (see Figs~\ref{D_W} and \ref{M_D}). Even when, following 
Chen et al. (1998), we select galaxies-absorption pairs which are 
likely to be physical association (since we couldn't apply the 
criterium on the galaxy-absorber correlation fonction, we only 
selected galaxies with impact parameter smaller than $500\h50$~kpc), 
we do not find any correlation between impact parameter and 
\lya\ rest equivalent width. This does not support a galactic halo 
model with radii as large as  $320\h50$~kpc as suggested by LBTW 
(using $q_0 = 0$ and $z\sim 0.3$).

\begin{figure}
\centerline{\psfig{figure=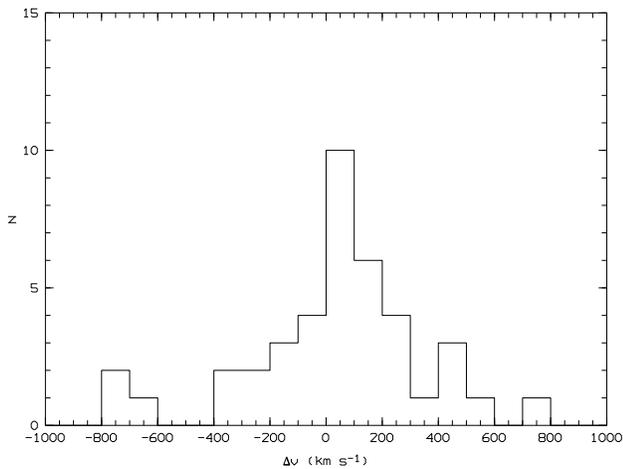,height=6.5cm,clip=t,angle=-90}}
\caption{\label{dv} Histogram of the \lya\ absorber-galaxy relative 
velocity}
\end{figure}

The relative velocity distribution of the \lya\ absorber-galaxy  associations
is shown in Fig.~\ref{dv}. The total sample of 39 associations is 25\% larger 
than that of LBB. This distribution remains narrow with a $\mathrm{HWHM}\simeq
100$\kms\ and 
the absolute relative velocity is smaller than 300\kms\ in 73\% of the cases.
The redshift measurement error, which corresponds to 120\kms , has no
influence on the shape of this distribution.
These results are very similar to those previously determined by LBB. They are 
also consistent 
with those derived from a survey of local \lya\ absorber-galaxy associations 
by Bowen et al. (1996),  the only difference being that the latter authors do 
not observe a peak around $\Delta v=0$\kms, which might be due to the smaller 
size of their sample.
Among the 14 galaxies without associated \lya\ absorption, the smallest value 
of  $\left|\Delta v\right|$ is  950\kms\ (the $z=0.3509$ galaxy in the field 
of PKS~1354+19) while for all the other 13 cases  $\left|\Delta v\right|\geq 
1400$\kms.
The galaxies without associated \lya\ absorption can therefore be easily 
separated from the sample of absorber-galaxy associations. Furthermore, the  
relative velocity distribution  between a galaxy and the nearest \lya\ 
absorber is  highly peaked for the \lya\ absorber-galaxy associations
 whereas it is flat for the non-associations. This strengthens the
 suggestion that, even if the absorbers are not directly linked to galaxies,
they are not distributed at random with respect to large-scale structures as 
traced by galaxies since the  $\Delta v$ distribution strongly departs from 
uniformity. 
\subsection{Association as a function of impact parameter}
\begin{figure}
\centerline{\psfig{figure=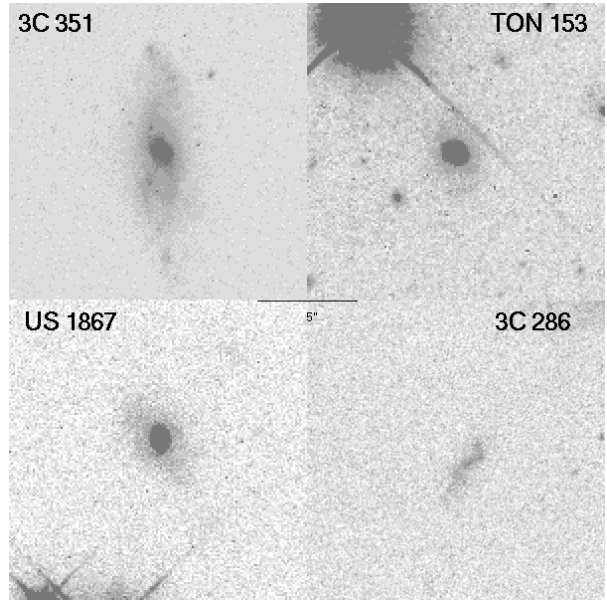,height=8cm,clip=t}}
\caption{\label{abs-imag} HST/WFPC2 images of four of the five known \lya\ 
absorbing galaxies}
\end{figure}

\begin{figure}
\centerline{\psfig{figure=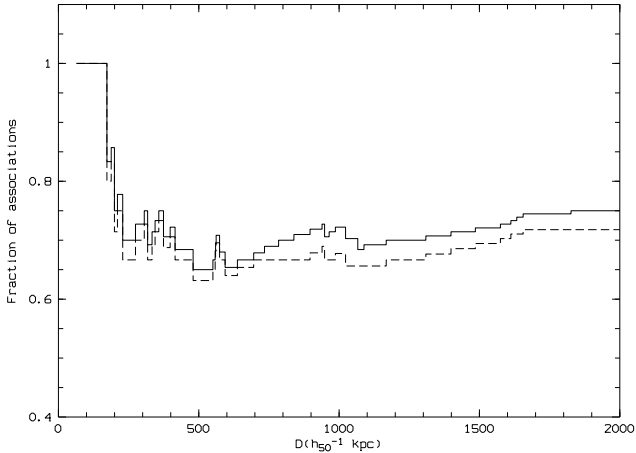,height=6.5cm,clip=t,angle=-90}}
\caption[]{\label{frac} Cumulative fraction of \lya\ absorbing galaxies 
versus impact parameter. The solid line refers to the present sample and the 
dotted one to that of LBB}
\end{figure}

The cumulated fraction of \lya\ absorber-galaxy associations  as a function of 
impact parameter is shown in Fig.~\ref{frac}. As in LBB, this fraction is
unity for impact parameters smaller that $175\h50$~kpc, which
suggests that  all field galaxies are most probably surrounded by huge 
($R\simeq 200\h50$~kpc) gaseous halos, whose external parts
give rise to \lya-only absorption (or more specifically with 
$\Wr(\mbox{\civ}1548)<$0.15~\AA\ at the 3$\sigma$ detection level). These 
galaxies are really \lya\ absorbing galaxies, in the sense that the absorbing 
is gravitationally or dynamically linked to them. 
There are five such galaxies in our sample, which characteristics are shown
in Table~\ref{absorbers}. HST/WFPC2 images of 4 of these 5 galaxies are 
available in the HST Science Archives, and we show the images of these 
objects in Fig.~\ref{abs-imag}. Two of them, in the fields of 3C~351 and 
US~1867, are also presented in Chen et al. (1998). As can be seen, these 
galaxies cover various morphological and/or spectral types. Therefore, 
it apears that, contrary to \mgii\ absorbing galaxies, \lya\ absorbing 
galaxies do not represent a specific sub-sample of galaxies, and that 
the presence of a very large ($R\sim 200\h50$~kpc) gaseous halo is a 
common characteritic of all classes of galaxies, maybe with the
possible exception of the ellipticals. 
\begin{table*}
\caption{\label{absorbers} Characteristics of the five \lya\ absorbing galaxies 
}
\begin{tabular}{rrrrrr}
\hline\noalign{\smallskip}
Field & PI Name & HST Dataset root& $\zg$ & Spectral features & morpholgy \\
QSO   &  PID    &                 &       &                   &           \\
\hline\noalign{\smallskip}
3C 286 & Bergeron  & U2B10A03T to & 0.3338 & \oii, \oiii, H$\alpha$ & diffuse irregular LSB \\
       &  5351     & U2B10A06T    &        &  and H$\beta$ emission &   \\
TON 153 & Steidel  & U2OM0901T to & 0.6715 & Ca\,{\sc ii} H \& K & S0 \\
        &   5984   & U2OM0904T    &        & and G band absorption  & \\
3C 351  & Disney   & U2WK0501T to & 0.0910 & \oiii, H$\alpha$, N\,{\sc ii} & Inclined spiral (Chen et al. 1997) \\
        & 6303     & U2WK0504T    &        & and S\,{\sc ii}  emission & \\
US 1867 & Lanzetta & U2X30301T to & 0.4435 & Ca\,{\sc ii} H \& K & S0 (Chen et al. 1997) \\
        & 5949     & U2X30303T    &        & and Fe absrption & \\
H 1821+64 &        &              & 0.2251 & Na\,{\sc i} absorption, & \\
          &        &              &        & H$\alpha$ emission \\
\hline\noalign{\smallskip}
\end{tabular}
\end{table*}

At intermediate impact parameter, the fraction of galaxies that are 
associated with a \lya\ absorbing cloud equals 60\% between 200 and 
$600\h50$~kpc, thus higher 
than the value of 40\% derived by Bowen et al. (1996) for lower 
redshift objects and $200<D<400\h50$~kpc, and in contradiction with 
the statement of Chen et al. (1997) that galaxies at impact parameter 
greater than $320\h50$~kpc are ``{\it almost never}" associated with a 
\lya\ absorption line. 
Both associated and non-associated galaxies cover a wide range of
spectral types (see Chen et al. (1998) for a display of a large number 
of both absorption-associated and non associated galaxies). The galaxy 
spectra (from our observations and those of LBTW) show either strong 
emission lines, \oii 3727 and H$\alpha$ and also in some cases H$\beta$
and \oiii, or the Balmer line series in absorption, or a strong \caii\ 
break. Consequently, neither the spectroscopic nor morphological type 
of a galaxy is a characteristic parameter of galaxies associated with 
\lya\ absorption.

At very large impact parameters, above $600\h50$~kpc,  there is an important 
fraction of associations (about 80\%), slightly higher for this new sample 
than previously found (Fig.~\ref{frac}). However, at these large impact
parameters, a large fraction of the associations could be fortuitious, since
these distances are close to the average separation between two galaxies. On
the contrary, the associations with projected separations below $600\h50$~kpc
are more likely to have a physical origin, depending on the characteristics of
the galaxies (brightness, class, environment ...), but the size of the
sample is still too small to determine the governing factors in these
associations (see a discussion on clustering in Sect.~\ref{groups}). 

These results are consistent with the recent analysis by  Dinshaw et al. (1997) 
of the properties of the absorption lines in a quasar close pair. The equivalent
width distribution of their sample of coincident and anticoincident \lya\ 
systems is best reproduced by models of flatten structures of very large 
dimensions ($r\sim 0.5\h50$~Mpc in the assumption of randomly inclined disks). 
This result is also in agreement with recent cosmological simulations which 
suggest that \lya\ absorbers trace large-scale filaments or sheets (Hernquist 
et al. 1996, Reidiger et al. 1998)
\subsection{\label{groups}Relation with groups or clusters of galaxies}
Despite the small size of the sample, we can  start to examine
the role of the galaxy environment as a factor for the
association with \lya\ absorption lines. There are five groups of 
galaxies (i.e. comprising from at least 2 to 4
members, with  velocities spanning less than 500\kms ), which comprise 
11 of the 53 galaxies of the sample. They all show associated \lya\ 
absorption. Among these five groups, two  in the field around TON~153 
(LBB) are part of a larger structure, with  other galaxies and \lya\ absorption lines at 
$\Delta v < 1500$\kms. The three other groups, all detected in the field of 
3C~286, are associated with a single isolated \lya\ line (see 
Table~\ref{cluster}). For three additionnal groups, one at $z=0.325$ in the 
field of 3C~95 (LBTW), and two at $z=0.07$ and 0.09 in the field of 3C~351 
(LBB), no meaningful limit 
($w_\mathrm{r,lim}$ of only about 1~\AA) could be set on the  corresponding 
\lya\ line.  Thus, all small groups of galaxies, for which there are
adequate spectroscopic data of the background quasars, have an associated 
\lya\ absorption line. The presence of these groups is not accounted for by
Chen et al. (1997) in their hypothesis of a $R(L^\star)\simeq500\h50$~kpc halo, which
would then contain one or more galaxy of the group it belongs to. 
Thus, we favour the existence of a smaller halo ($R \simeq 200\h50$
~kpc), and absorption by gas belonging to the structures (either large scale 
structures of groups/clusters) traced by galaxies. 

The known number of members of theses groups is of course a lower limit to 
their actual membership, and only a complete spectrocopic survey will
reveal the nature of these associations. 
We have performed Monte-Carlo simulations to derive the probability to obtain 
the observed velocity separation for 2 or 3 galaxies in a cluster or 
group of given velocity dispersion $\sigma_v$, which show that the $z=0.490$ 
group does not trace a rich cluster with $\sigma_v \ge 1400$\kms , at 
a $3\sigma$ confidence level. 

\begin{figure}
\centerline{\psfig{figure=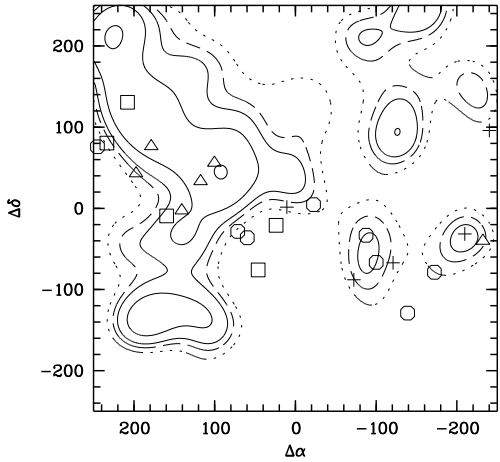,height=6.5cm,clip=t,angle=-90}}
\caption[]{\label{density} Galaxy density map of the field aroun
3C~286. The
dotted line is $1\sigma$ (Poisson statistics) below the 'canonical' 
average density, dashed line is $2\sigma$ above this value, solid 
lines are 4,8 and $16\sigma$ above this value. Crosses: $z=0.33$ group,
triangle: galaxies in the range $[0.4:0.5]$, squares: galaxies in the 
range $[0.2:0.25]$, circles: other redshifts}
\end{figure}

In the field of 3C~286, the actual total number of galaxies with 
apparent magnitude $\mr \le 23$ is 396. The average number of galaxies 
expected in a $500\arcsec \times 500\arcsec$ field is 350, as derived 
from the differential
magnitude-number counts of field galaxies given by Le~Brun et al. (1993) : 
\begin{equation}
\label{dnds}
{\partial^2 N\over \partial S \partial m} = 10^{\left(0.366\pm 0.01\right)\mr -
  (4.24\pm0.31)},
\end{equation}
 in very good agreement with those obtained by  Metcalfe et al. (1991), Smail 
et al. (1995) or Driver et al. (1994). 
Thus, the field around 3C~286
appears to be slightly denser than average ($2.5\sigma$ effect). However,
since the physical size of the field is greater than $3\h50$~Mpc at the
redshifts of the \lya\ absorbers, the presence of large clusters (similar to
that surrounding H~1821+64, see LBB) containing
more than 100 galaxies is excluded up to $z=0.5$, which includes all 
the \lya\ absorbers associated with groups of galaxies. 
Using the same kind of algorithm as in Le Brun et al.
(1993), we have drawn the galaxy density map of the field. As can be 
seen in Fig.~\ref{density}, the field is highly inhomogeneous, with 
two peaks at more than 8$\sigma$ above the average density. These peaks are all located eastwards to the
quasar, as well as most of the members of the
$z=0.248$  group and most of the galaxies in the redshift range $[0.4,0.5]$
(triangles and squares on Fig.~\ref{density}).  
These structures comprise a few tens of galaxies, and are likely to be groups or
small clusters, their size being about $400\h50$~kpc at $z=0.5$. 
The presence or absence of associated 
metal line of different ionization levels could allow to study the 
ionization conditions that prevail in these absorbers, as well as
their average metallicity.
\begin{table}
\caption{\label{cluster}Clustering of the \lya\ absorbers in the field of
  3C~286} 
\begin{tabular}{llllll}
\noalign{\smallskip}\hline \\
$\za$ & $N_\mathrm{g}$ & $\zg$ & Size & $\MAB$ & 
$\left|\Delta v_\mathrm{g-g}\right|$ \\
      &                &       & $h_{50}^{-2}$~kpc$\times$kpc& & 
\kms \\
\noalign{\smallskip}\hline \\
 0.3348 & 2 & 0.3316 & $800\times450$   & $-19.58$ &  500 \\
        &   & 0.3338 &                  & $-19.95$ &      \\
 0.4309 & 2 & 0.4334 & $600\times 150$  & $-21.45$ &  330 \\
        &   & 0.4350 &                  & $-21.24$ &      \\
 0.4917 & 3 & 0.4895 & $3450\times650$  & $-21.43$ &  100 \\
        &   & 0.4900 &                  & $-22.61$ &      \\
        &   & 0.4900 &                  & $-19.71$ &      \\
\noalign{\smallskip}\hline \\
\end{tabular}
\end{table}

There is at present only one case of an identified galaxy belonging to a 
rich cluster not associated with the quasar : 
in the field of 3C~273, a galaxy is a member of  the   
rich cluster, A~1564 at $z=0.078$ with 54 identified galaxies (see Morris et 
al. 1993 for a complete discussion), but no associated \lya\ absorption is 
detected. This could be explained by the physical conditions which prevail in
such a dense environment: the interstellar gas is stripped from individual 
galaxies by tidal effects and the intracluster gas is too highly ionized to 
contain detectable amounts of neutral hydrogen. 

One would naturally expect that \lya\ absorbers are also found near
closely interacting galaxies, for which gas is  expected to be ejected 
at very large distance by tidal effects. Our survey is
at a too early stage (only one such case observed so far) to estimate the 
gaseous cross-section of interacting galaxies.

\end{document}